\documentclass[12pt]{article}
\usepackage{cite,color}

\makeatletter
\@addtoreset{equation}{section}

\makeatletter
\renewcommand\section{\@startsection {section}{1}{\z@}%
                                   {-3.5ex \@plus -1ex \@minus -.2ex}%nn
                                   {2.3ex \@plus.2ex}%
                                   {\normalfont\large\bfseries}}
\renewcommand\subsection{\@startsection{subsection}{2}{\z@}%
                                     {-3.25ex\@plus -1ex \@minus -.2ex}%
                                     {1.5ex \@plus .2ex}%
                                     {\normalfont\bfseries}}

\def\baselinestretch{1.2}
\newcommand{\be}{\begin{equation}}
\newcommand{\ee}{\end{equation}}
\newcommand{\beq}{\begin{eqnarray}}
\newcommand{\eeq}{\end{eqnarray}}
\newcommand{\oo}{\"o}

\def\[{\left [}
\def\]{\right ]}
\def\({\left (}
\def\){\right )}

\def\S{{\bf S}}

\def\CP{{\cal P}}

\def\l{\ell}
\def\j{\delta}

\begin{document}
\begin{titlepage}

\begin{flushright}
hep-th/0306131\\
SU-ITP-03/15\\
UCB-PTH-03/11\\
LBNL-52960
\end{flushright}
%\vspace{12 mm}

\vfil\
%vfil

\begin{center}

{\Large{\bf Black strings in asymptotically plane wave geometries}}
\vfil

\vspace{3mm}

Eric G. Gimon$^{a}$, Akikazu Hashimoto$^a$, Veronika E. Hubeny$^b$,\\
\vspace{1mm}
Oleg Lunin$^a$ and  Mukund Rangamani$^c$ \\

\vspace{8mm}

$^a$ Institute for  Advanced Study, %School of Natural Sciences,
Einstein Drive, Princeton, NJ 08540\\

$^b$ Department of Physics, Stanford University, Stanford, CA 94305\\

$^c$ Department of Physics, University of California,
Berkeley, CA 94720\\

$^c$ Theoretical Physics Group, LBNL, Berkeley, CA 94720

\vfil

\end{center}

%%%%%%%%%%%%%%%%%%%%%%%%%%%%%%%%%%%%%%%%%%%%%%%%%%%%%%%%%%%%%%%%%%%%%%%%%%%%%%%%%%%%%%%
\begin{abstract}
\noindent
We present a class of black string spacetimes which asymptote to
maximally symmetric plane wave geometries. Our construction will
rely on a solution generating technique, the {\it null Melvin
twist}, which deforms an asymptotically flat black string spacetime
to an asymptotically plane wave black string spacetime while
preserving the event horizon.
\end{abstract}
%%%%%%%%%%%%%%%%%%%%%%%%%%%%%%%%%%%%%%%%%%%%%%%%%%%%%%%%%%%%%%%%%%%%%%%%%%%%%%%%%%%%%%%%%
\vspace{0.5in}

\end{titlepage}
\renewcommand{\baselinestretch}{1.05}  %Line spacing
%%%%%%%%%%%%%%%%%%%%%%%%%%%%%%%%%%%%%%%%%%%%%%%%%%%%%%%%%%%%%%%%%%%%%%%%%%%%%%%%%%%%%%%%%%%%%

\section{Introduction}

Maximally symmetric plane waves have emerged as an important
background space-time of string theory \cite{BFHP}.  Just like
Minkowski, anti de Sitter, and de Sitter spaces, maximally symmetric
plane wave geometries are homogeneous.  Although all curvature
invariants vanish, they have causal structure different from that of
flat space \cite{BN,MaRoa,HRcausal, Marob} and are not globally
hyperbolic \cite{Pghyper}. The most intriguing aspect of certain
maximally symmetric plane waves is the fact that they admit dual field
theory description through the correspondence of \cite{BMN}.

Schwarzschild black holes provide important insights into the nature
of gravity.  While black objects in other maximally symmetric spaces
are well known, analogous solutions in plane wave geometries have yet
to be constructed. Nonetheless, it was shown in \cite{HRnoho} that
global null Killing isometry is consistent with the existence of an
extremal event horizon.  Indeed, extremal vacuum plane wave black hole
solutions with regular horizon were constructed in \cite{HRnullsol}.
Various extremal and non-extremal deformations of the maximally plane
wave geometry were also considered in \cite{Herdeiro:2002ft,LPV}.
More recently, pure Schwarzschild black string deformation of the six
dimensional plane wave geometry with regular event horizon was
constructed in \cite{GiHa}.  Unfortunately, that Schwarzschild
deformation has the effect of distorting the asymptotic geometry by a
finite amount and can not be considered as a black string in an
asymptotically plane wave geometry.

In this article, we present a systematic solution generating technique
which can be used to construct a general class of asymptotically plane
wave solutions.  Our method is based on the observation of \cite{AlGo}
that certain class of plane wave geometries can be generated by
applying a sequence of manipulations, which we will call the {\it null
Melvin twist}, to Minkowski space.  By applying the same sequence of
manipulations starting from a black string solution, we are able to
generate a large class of black string deformations of plane wave
geometries with a regular horizon.  These solutions are characterized
by the mass density of the black string and the scale of the plane
wave geometry in the rest frame of the black string.  For dimensions
greater than six, the deformation due to the presence of the black
string decays at large distances in the directions transverse to the
string. Hence in these cases, our solutions describe geometries with
(radially) plane wave asymptotics. One can also construct black hole
solutions in asymptotically G\"odel universes \cite{GGHPR} by further
dualizing these solutions \cite{Herdeiro:2002ft,BGHV,HaTa}.

Since the asymptotically plane wave black string solutions we find in
this paper have a regular horizon, it is straightforward to compute
its area.  We find that this area is identical to the area of the
black string before the null Melvin twist.  We will provide a proof
that the null Melvin twist is a procedure which keeps the area of the
horizon invariant.

The null Melvin twist can be used to generate a plane wave background
with the same metric as the Penrose limit of $AdS_5 \times \S^5$ in
type IIB supergravity which was identified in \cite{BMN} as admitting
a dual field theory description. Unfortunately, these two backgrounds
differ from each other in their matter field content. As such, this
solution generated using a null Melvin twist does not yet have an
obvious dual field theory interpretation. Nonetheless, the null Melvin
twist yields an explicit black string solution which should provide
useful hints for finding the black string deformation of the plane
wave of \cite{BMN}.

The organization of the paper is as follows. In section 2, we
construct the black string solution in ten dimensional asymptotically
plane wave background and describe some of its properties.  In section
3, we describe various generalizations including adding charges and
angular momenta.  In section 4, we describe the analogous construction
for plane wave black strings in other dimensions.  Concluding remarks
are presented in section 5.

%%%%%%%%%%%%%%%%%%%%%%%%%%%%%%%%%%%%%%%%%%%%%%%%%%%%%%%%%%%%%%%%%%%%%%%%%%
\section{Asymptotically plane wave black string in 10 dimensions}
%%%%%%%%%%%%%%%%%%%%%%%%%%%%%%%%%%%%%%%%%%%%%%%%%%%%%%%%%%%%%%%%%%%%%%%%%%

In this section, we describe the black string in an asymptotically 10
dimensional plane wave geometry. We will denote the maximally
symmetric plane wave geometry in $d$ dimensions by $\CP_d$. The metric
of $\CP_{10}$ is
\beq ds^2 &=&
-dt^2 + dy^2 - \beta^2
\sum_{i = 1}^8 \, x_i^2 \, (dt + dy)^2 + \sum_{i=1}^8 \,  dx_i \,
dx^i \  .
\label{ppten}
\eeq
If this metric is supported by a self dual 5-form field strength as in
\cite{BFHP}, this background is a maximally supersymmetric solution of
type IIB supergravity with isometry group $SO(4)\times SO(4)$.  In
fact, there exists a one-parameter family of metrics \cite{BeRo,
Michelson} with the same spacetime geometry, but with the metric
supported by a combination of the RR 5-form and a RR 3-form of IIB
supergravity.  A generic member of this family preserves 28
supercharges and a $U(2)\times U(2)$ isometry group. Another special
member of this one parameter family is a solution which is supported
by the RR 3-form alone, with the isometry group enhanced to $U(4)$.
Using S-duality, we then obtain a new $\CP_{10}$ solution which is
supported entirely by fields in the NS-NS sector.

The $\CP_{10}$ solution with NS-NS fields above is particularly well
suited as a background for introducing a black string for two
reasons. First, the action of the isometry group $U(4)$ is transitive
on the transverse seven-sphere (the orbits are $U(4)/U(3)$), which
means that we can expect to modify the metric with functions of the
transverse distance only and thus can use brute force to find the
black string solution.  Second, and more importantly, we can find the
black string by using the fact that Minkowski space is related this
$\CP_{10}$ by a null Melvin twist.

%%%%%%%%%%%%%%%%%%%%%%%%%%%%%%%%%%%%%%%%%%%%%%%%%%%%%%%%%%%%%%%%%%%%%%%%%%%%%
\subsection{Null Melvin twist}
%%%%%%%%%%%%%%%%%%%%%%%%%%%%%%%%%%%%%%%%%%%%%%%%%%%%%%%%%%%%%%%%%%%%%%%%%%%%%

In this subsection, we will describe the sequence of solution
generating manipulations which we call the null Melvin twist. We start
by constructing a neutral black string in $\CP_{10}$ as an example. As
was shown in \cite{AlGo}, these same manipulations applied to
Minkowski space generate $\CP_{10}$. As the construction involves only
the NS-NS sector fields, they can be applied to any supergravity with
an NS-NS sector. To be specific, let us describe the case where we
start with a type IIB theory. To facilitate the duality transformations,
we will write the metric in string frame.

\begin{enumerate}

\item 
Consider a Schwarzschild black hole solution in 8+1 dimensions
\cite{tangherlini} embedded into type IIB supergravity
\be  ds_{str}^2  =  -f(r) \, dt^2 + dy^2 + {1 \over f(r)} \, dr^2 + r^2 \, d
\Omega_7^2 \ . \label{vacuumbs} \ee
\be f(r)  =  1 - {M \over r^6} \ .
\label{schwfn}
\ee
which describes a black string with mass density $M$. This solution is
translationally invariant along $y$.

\item \label{step2} Boost the geometry in the $y$ direction by an amount
$\gamma$. This gives rise to a black string with net momentum $P_y = M
\sinh \gamma \cosh \gamma$.

\item  T-dualize along $y$. This gives a  solution of IIA
supergravity with fundamental string charge $Q_{F1} = M \sinh \gamma \cosh \gamma$.
Translation along $y$ and $SO(8)$ rotations along the transverse
$\S^7$ remain isometries of this geometry.

\item Twist the rotation of $\S^7$ along $y$. By twisting, we mean
parameterizing the plane transverse to the string in cartesian
coordinates $x_i$, and making the following change of coordinates
\beq
x_1 + i x_2 &\rightarrow & e^{i \, \alpha \,  y} (x_1 + i x_2) \cr
x_3 + i x_4 &\rightarrow & e^{i \, \alpha \, y} (x_3 + i x_4) \cr
x_5 + i x_6 &\rightarrow & e^{i \, \alpha \, y} (x_5 + i x_6) \cr
x_7 + i x_8 &\rightarrow & e^{i \, \alpha \, y} (x_7 + i x_8)  \ .
\eeq
where the parameter $\alpha$ controls the amount of twisting. For
the sake of simplicity, we twist in all four planes of rotation by
the same amount. This has the effect of replacing the metric on
the 7-sphere according to
\be d  \Omega_7^2 \rightarrow 
d \Omega_7^2 +  \alpha \, \sigma \, dy +  \alpha^2 \,
 dy^2 \ee
where
\be {r^2 \, \sigma \over 2} =
x_1 dx_2 - x_2 dx_1 +
x_3 dx_4 - x_4 dx_3 +
x_5 dx_6 - x_6 dx_5 +
x_7 dx_8 - x_8 dx_7 \ .
\ee

\item T-dualize along $y$.  This geometry now corresponds to a black
string in type IIB supergravity with momentum $P_y = M \sinh \gamma
\cosh \gamma $ sitting at the origin of the Melvin universe \cite{Melvin:1964qx}, and has a $U(4)$ isometry
group.

\item Boost the solution by $-\gamma$  along $y$. The purpose of this
boost is to cancel the net momentum due to the original boost 
performed at step \ref{step2}.

\item  Now, we perform a double scaling limit, wherein the boost
$\gamma$ is scaled to infinity and the twist $\alpha$ to zero keeping
\be  \beta  = {1 \over 2} \, \alpha \, e^\gamma = \mbox{fixed} \ .
\label{doublescaling}
\ee

The end result is the black string in an asymptotically $\CP_{10}$ spacetime.
\beq
ds_{str}^2 & =& - {f(r)\, \left(1 + \beta^2 \, r^2
\right) \over k(r)} \, dt^2 -  \, {2 \, \beta^2 \, r^2 \, f(r) \over k(r)} \,
dt \, dy + \left( 1  -{ \beta^2\,  r^2 \over k(r)} \right) \, dy^2  \cr
& & \qquad  + {dr^2 \over f(r)} + r^2 \, d\Omega_7^2 - {\beta^2 \, r^4 
\, (1 - f(r))
\over 4 \, k(r)}\, \sigma^2  \cr
e^{\varphi} &=& { 1 \over \sqrt{k(r)}} \cr
B & = &{\beta \, r^2 \over 2 k(r)}\, \left(f(r) \, dt + dy\right) \wedge
\sigma \label{tendbstr}  \eeq
where
\be k(r) = 1 + {\beta^2 M  \over r^4} \ .
\label{kdeftend}
\ee
\end{enumerate}

Steps 3 through 5 are the manipulations involved in generating an
ordinary Melvin flux tube solution by twisting along a spatial
isometry direction $y$ which was originally described in
\cite{Dowker:1994bt}. Steps 2 and 6 boost the direction in the $t$-$y$
plane along which the Melvin twist is performed.  The final step has
the effect of scaling the isometry direction of the Melvin twist to be
null.  It is therefore natural to refer to this sequence of steps as
the null Melvin twist.

%%%%%%%%%%%%%%%%%%%%%%%%%%%%%%%%%%%%%%%%%%%%%%%%%%%%%%%%%%%%%%%%%%%%%%%%%%%%%%%%%%
\subsection{Properties of the asymptotically $\CP_{10}$ black string solution}
%%%%%%%%%%%%%%%%%%%%%%%%%%%%%%%%%%%%%%%%%%%%%%%%%%%%%%%%%%%%%%%%%%%%%%%%%%%%%%%%%%

The solution (\ref{tendbstr}) is very simple. By inspection, if we
set $M$ to zero the solution reduces to the $\CP_{10}$ geometry
described at the beginning of this section.  On the other hand,
setting $\beta=0$ will reduce the solution to the black string
solution (\ref{vacuumbs}).  There is a regular horizon at
\be r_H = M^{1/6} \ee
which persists for finite values of $\beta$.  One can therefore
interpret (\ref{tendbstr}) as the black string deformation of
$\CP_{10}$.  Furthermore, since both $f(r)$ and $k(r)$ asymptote to 1
as $r$ is taken to be large, the effect of $M$ decays at large
$r$. Unlike the six dimensional solution described in \cite{GiHa}
which deformed the geometry by a finite amount at large $r$,
(\ref{tendbstr}) is a black string solution which genuinely asymptotes
to $\CP_{10}$.

Because both the dilaton and the metric asymptote to $\CP_{10}$ in
(\ref{tendbstr}), one can unambiguously define the area of the
horizon in Einstein frame with the same asymptotics. The area of
the fixed ($r$,$t$)-surface in Einstein frame for (\ref{tendbstr})
is
\be
{\cal A} =   L \, \sqrt{k(r) - \beta^2 r^2}  \;   r^7 \, \Omega_7 \ 
\ee
where $\Omega_7 = {\pi^4 / 6}$ is volume of a unit $\S^7$ and $L$ is
the length of the translationally invariant $y$-direction.  At the
horizon, this evaluates to
\be
{\cal A}_H = L \, M^{{7/6}} \, \Omega_7 \ . \ee
It is tempting to interpret this area in Planck units
\be
S  = { L\, M^{{7 / 6}} \, \Omega_7 \over 4 \, G_{10}}
\label{bfhpentropy}
\ee
as an entropy of some sort. A remarkable fact is that this quantity is 
independent of the parameter $\beta$.

It would also be interesting to compute the temperature associated to
this black string.  Ordinarily one computes the temperature of a black
object in terms of the surface gravity
\be
\kappa^2 = - {1 \over 2  } \, \left( \nabla^a \xi^b \right)
\left(\nabla_a \xi_b \right),
\label{surfgrav}
\ee
where $\xi^a$ is the null generator of the horizon.  The
temperature is then given in terms of the surface gravity as $ T_H =
\kappa/2 \pi$.

For the solution (\ref{tendbstr}), the null generator of the horizon is 
simply the Killing vector 
\be \xi^a = \left({\partial \over \partial t} \right)^a \ . 
 \label{killingV} \ee
Special care is necessary with regards to the normalization of this
Killing vector in the computation of the temperature.  In
asymptotically flat space, for example, it is natural to normalize the
Killing vector so that it is of unit norm asymptotically. It is a
priori not clear what constitutes a natural normalization of the
Killing vector in an asymptotically plane wave geometry. Let us
therefore use a normalization such that the Killing vector takes
precisely the form (\ref{killingV}) and interpret the temperature as
being measured in units of inverse coordinate time $t$. With this
caveat in mind, the temperature at the horizon of the solution
(\ref{tendbstr}) is found to be
\be T_H = {3 \over 2 \pi} M^{-1/6} \ .\ee
What is remarkable about these results is the fact that both the
temperature and entropy are independent of the parameter $\beta$.  We
will in fact show in the appendix that the area of the horizon is
invariant under a null Melvin twist for a general class of spacetimes.
This would appear to indicate that the parameter $M$ has a natural
interpretation as the mass density of the black string. This is a
rather non-trivial statement since a proper notion of mass analogous
to the ADM mass for an asymptotically plane wave geometry has not yet
been defined. 

The solution (\ref{tendbstr}) is free of closed
time-like curves so long as the $y$ coordinate is decompactified at
the end of chain of dualities.  If the $y$ coordinate is compact,
closed time-like curves can appear just as in the case of ordinary
plane-waves.

%%%%%%%%%%%%%%%%%%%%%%%%%%%%%%%%%%%%%%%%%%%%%%%%%%%%%%%%%%%%%%%%%%%%%%%%%%%%
\section{Generalizations}
%%%%%%%%%%%%%%%%%%%%%%%%%%%%%%%%%%%%%%%%%%%%%%%%%%%%%%%%%%%%%%%%%%%%%%%%%

The null Melvin twist construction is extremely simple and can be
applied to generate a wide variety of asymptotically plane wave
geometries.  In this section we will describe a few examples.

%%%%%%%%%%%%%%%%%%%%%%%%%%%%%%%%%%%%%%%%%%%%%%%%%%%%%%%%%%%%%%%%%%%%%%%%%%
\subsection{Rotating black holes}
%%%%%%%%%%%%%%%%%%%%%%%%%%%%%%%%%%%%%%%%%%%%%%%%%%%%%%%%%%%%%%%%%%%%%%%%%%%
One simple generalization of (\ref{tendbstr}) is to add angular momentum.
Consider a rotating black string in type IIB supergravity  \cite{Myers:1986un}.
\be ds_{str}^2 = - dt^2 + \(1 -f(r) \) \, \( dt + {\l \over 2} \sigma
\)^2 + dy^2 + {1 \over h(r)} dr^2 + r^2 d \Omega_7^2 \ . \ee
\label{Kerrbs}
where
\be
f(r) = 1 - {M \over r^6} \qquad {\rm and} \qquad
h(r) = 1 - {M \over r^6} + {M \l^2 \over r^8} \ .
\ee
In general, rotating black strings in 10-dimensions admit four
independent angular momentum charges.  We have taken all four
angular momenta to equal $\l$ for simplicity.

Applying the null Melvin twist procedure of the previous section
leads to the following solution of type IIB supergravity
\beq  ds_{str}^2 & =& - {f(r) + \beta^2 r^2 \, h(r) \over k(r)} \,
dt^2 - \, {2 \beta^2 r^2 h(r) \over k(r)} \, dt \, dy + \left( {1 -
\beta^2 r^2\, h(r) \over k(r)} \right) \, dy^2  \cr &&
\qquad  \qquad + {dr^2 \over
h(r)} + r^2 \, d\Omega_7^2  - {M \over 4 \, r^2 \, k(r) }\,\left(\beta^2 - {\l^2
\over r^4}\right)\, \sigma^2 - {M \l \over r^6
k(r)}\,\sigma \, dt \cr \cr  e^{\varphi} &=& { 1 \over
\sqrt{k(r)}} \cr  B & = &{\beta r^2 \over 2 k(r)}\, \left(h(r)
\,dt\wedge \sigma +  (1 + {M \l^2 \over r^8})\, dy\wedge
\sigma + {2 M \l \over r^8} \,dt\wedge dy\right)
\label{ppspinbstr}
\eeq
where as before
\be
k(r) = 1 + {M \beta^2 \over r^4} 
\ee
The inner and outer horizons are located at the roots of $h(r)$.
Just as in the non-rotating case, this geometry asymptotes to
plane wave for large $r$ or small $M$, but reduces to a rotating
black string in the small $\beta$ limit.  The horizon area and the
surface gravity are independent of $\beta$ and agree with the
results for a rotating black string in asymptotically flat space.
The outer horizon of the $\l \ne 0$ solution carries non-vanishing
angular velocity
\be \omega_H = -{2 \l \over r_H^2} \ . \ee

%%%%%%%%%%%%%%%%%%%%%%%%%%%%%%%%%%%%%%%%%%%%%%%%%%%%%%%%%%%%%%%%%%%%%%%%%%%%%%
\subsection{Charged black strings}
%%%%%%%%%%%%%%%%%%%%%%%%%%%%%%%%%%%%%%%%%%%%%%%%%%%%%%%%%%%%%%%%%%%%%%%%%%%%

It is also easy to add charges to (\ref{tendbstr}). Simply start with
the non-extremal fundamental string solution and apply the null Melvin
twist.  This gives
\beq
ds_{str}^2 &=& 
- {f(r) \over k_\delta(r)} \, dt^2- {\beta^2 \, r^2  \over k_\j(r)}
\left(1-\frac{M\cosh^2\delta}{r^6}\right)
\,(dt+dy)^2+{1 \over k_\delta(r)} dy^2  + 
{1 \over f(r)} \, dr^2 \cr 
&& \qquad  + r^2 \, d\Omega_7^2 +  
{\beta \, M \, \sinh 2\j \over 2\, r^4\,k_\j(r)}\, 
(dt+dy)\, \sigma -
{\beta^2 M \over 4\, r^2\, k_\j(r)} \, 
\sigma^2 \cr
B&=& {M\, \sinh 2 \j \over 2\, r^6\, k_\j(r)} 
\, dt\wedge dy
+{\beta \,r^2 \,\over 2 \, \, k_\j(r) } \,
\( f(r)  dt + dy \) \wedge \sigma \cr
e^{\varphi}&=& {1 \over \sqrt{k_\j(r)}} 
\label{ppfbstr}
\eeq
Here, we introduced
\be 
k_\delta(r) \equiv 1+\frac{M\sinh^2\delta}{r^6}+\frac{M\beta^2}{r^4} \ee
A non-extremal D-string solution can be obtained in the same way if we
perform the null Melvin twist starting from the nonextremal D-string
instead of the fundamental string. The extremal limit of this solution
is closely related to the solutions described in
\cite{Bain:2002nq,Cvetic:2002nh}.

%%%%%%%%%%%%%%%%%%%%%%%%%%%%%%%%%%%%%%%%%%%%%%%%%%%%%%%%%%%%%%%%%%%%%%%%%%%%%%
\subsection{General twists}
%%%%%%%%%%%%%%%%%%%%%%%%%%%%%%%%%%%%%%%%%%%%%%%%%%%%%%%%%%%%%%%%%%%%%%%%%%%%%

So far, we have considered the case where one twists the four complex
planes transverse to the black string equally.  One can readily
generalize this to independent twists in each of the four complex
planes
\beq
x_1 + i x_2 &\rightarrow & e^{i \alpha\,  v_1 \,  y} (x_1 + i x_2) \cr
x_3 + i x_4 &\rightarrow & e^{i \alpha\,  v_2 \,  y} (x_3 + i x_4) \cr
x_5 + i x_6 &\rightarrow & e^{i \alpha \, v_3 \, y} (x_5 + i x_6) \cr
x_7 + i x_8 &\rightarrow & e^{i \alpha \, v_4 \, y} (x_7 + i x_8)  \ . \label{gentwist}
\eeq

Following the chain of dualities one obtains
\beq
ds_{str}^2 & =& - {f(r) \,  \left(1 + \beta^2 \, r^2 \, |\sigma_v|^2
\right) \over k_v(r,\Omega_7)} \, dt^2 -  \, {2 \, \beta^2 \, r^2 \,
|\sigma_v|^2 \, f(r) \over k_v(r, \Omega_7)} \,dt \, dy +
\left( 1  -{ \beta^2\,  r^2 \, |\sigma_v|^2 \over
k_v(r,\Omega_7)} \right) \, dy^2  \cr
& & \qquad  + {dr^2 \over f(r)} + r^2 \, d\Omega_7^2 - {\beta^2 \,
 r^4
\, (1 - f(r))  \over 4 \, k_v(r,\Omega_7)}\, \sigma_v^2  \cr
e^{\varphi} &=& { 1 \over \sqrt{k_v(r,\Omega_7)}} \cr
B & = &{\beta \, r^2 \, \over 2 \, k_v(r,\Omega_7)}\, \left(f(r)
\, dt + dy\right) \wedge
\sigma_v
\eeq
with
\be
k_v(r,\Omega_7) = 1 + {\beta^2 M  |\sigma_v|^2\over r^4},
\ee
\beq  { r^2  \sigma_v \over 2} &=&
v_1\, (x_1 dx_2 - x_2 dx_1) +
v_2 \, (x_3 dx_4 - x_4 dx_3) \cr & +&
v_3 \, (x_5 dx_6 - x_6 dx_5) +
v_4 \, (x_7 dx_8 - x_8 dx_7)  \ , \eeq
and
\be 
|\sigma_v|^2 = {1 \over r^2} \left(
v_1^2 (x_1^2 + x_2^2) +
v_2^2 (x_3^2 + x_4^2) +
v_3^2 (x_5^2 + x_6^2) +
v_4^2 (x_7^2 + x_8^2)  \right) \ . \ee
Note that since the $U(4)$ isometry is broken, $k_v(r,\Omega_7)$ can
now depend non-trivially on the coordinates of $\S^7$. One can
explicitly check, though, that the horizon area and the surface
gravity remain independent of $\beta$ and the $v_i$'s.

To be completely general, one can construct a solution with 13
independent parameters $M$, $\beta$, $P_y$, $Q_{F1}$, $Q_{D1}$,
$\l_1$, $\l_2$, $\l_3$, $\l_4$, $v_1$, $v_2$, $v_3$, and $v_4$. We
will not write this most general solution explicitly.

%%%%%%%%%%%%%%%%%%%%%%%%%%%%%%%%%%%%%%%%%%%%%%%%%%%%%%%%%%%%%%%%%%%%%%%%%%%%%%%

\section{Asymptotically plane wave black string in other dimensions}
%%%%%%%%%%%%%%%%%%%%%%%%%%%%%%%%%%%%%%%%%%%%%%%%%%%%%%%%%%%%%%%%%%%%%%%%%%%%%%%

Another natural generalization of our procedure is to apply the null
Melvin twist to black branes smeared in more dimensions. A black
$p$-brane in 10 dimensions, when compactified along $p-1$ of the
translationally invariant directions, will look like a black string in
$d = 11-p$ dimensions.  The $p-1$ extra dimensions play a spectator
role, and so by applying the null Melvin twist on the effective $d$
dimensional black string, one can construct black string solutions
which are asymptotically $\CP_d$.

The metric of black $(11 - d)$-brane is simply
\beq
 ds_{str}^2 &=& -f_d(r)\, dt^2 + dy^2 + {1 \over f_d(r)} \, dr^2 + r^2 \, d
\Omega_{d-3} + \sum_{i=1}^{10-d} \, dz_i^2 \ , \cr
f_d(r) & =&  1 -
 {M \over
r^{d-4}}\ . \eeq
To simplify the discussion, let us restrict to even values of $d$.
Then there are $(d-2)/2$ independent null Melvin twist parameters
that one can independently adjust.  Let us further take all the 
$(d-2)/2$ twist parameters to be equal for simplicity. Then, we
find that the supergravity solution for the neutral black string
in $\CP_d$ takes the form
\beq  ds_{str}^2 & =& - {f_d(r)\left(1 + \beta^2 \, r^2 \right)
\over k_d(r)} \, dt^2 - {2 \, \beta^2 \, r^2 \, f_d(r) \over k_d(r)}
\, dt  \, dy +  \left( 1  -{
\beta^2 \, r^2 \over k_d(r)} \right) \, dy^2  \cr  && \qquad +
{dr^2 \over f_d(r)} + r^2 \, d\Omega_{d-3}^2 -  {r^4 \, \beta^2 \,
( 1- f_d(r)) \over 4 \,  k_d(r)}\, \sigma_d^2  + \sum_{i=1}^{10-d} dz_i^2 \cr
e^{ \varphi} &=& { 1 \over \sqrt{k_d(r)}} \cr
B & = &{\beta \, r^2 \over 2 \, k_d(r)}\, \left(f_d(r) \, dt +
dy\right) \wedge \sigma_d
\label{ppdschw}
\eeq
with
\be
k_d(r) = 1 + {   \beta^2\, M \over r^{d-6} }
\ee
For $d > 6$, (\ref{ppdschw}) asymptotes to $\CP_d$ and closely
resembles $\CP_{10}$ in many ways. 

The $d=6$, is special in that $k_6(r)$ does not asymptote to 1. This
is similar to conical deficits which arise as a result of mass
deformation in 2+1 dimensions and cause the background to be deformed
by a finite amount even for large $r$.  Nonetheless, (\ref{ppdschw})
for $d=6$ can be considered as a black string deformation of $\CP_6$
in the sense that in the small $M$ limit, the solution reduces to
$\CP_6$.  The black string deformation of $\CP_6$ was also constructed
in \cite{GiHa}, but was presented in a slightly different form.  Let
us compare our solution to the solution presented in \cite{GiHa} in
some more detail.

For $d=6$,  (\ref{ppdschw}) becomes 
\beq
ds_{str}^2 &=& -
{(1 - {M \over r^2}) (1 +\beta^2 \,r^2) \over 1 +
\beta^2 M} \, dt^2 -  {2\beta^2 r^2 (1 - { M\over r^2} ) \over 1 +
\beta^2 M} \, dt \, dy + \left( 1 - {\beta^2  r^2 \over 1 + \beta^2 M}
\right) \, dy^2 \cr  &&  \qquad + {dr^2 \over 1 -{M \over r^2} } + r^2
\, d\Omega_3^2 - {r^2
\over 4} \, \left({\beta^2 M  \over 1 + \beta^2 M} \right)
\, \sigma_6^2 \cr
e^\varphi & = & {1 \over \sqrt{1 + \beta^2 M}} \cr
B & = &{\beta \, r^2 \over 2 \, (1 + \beta^2 M)}\, \left( \left(1 - {M \over r^2} \right) \, dt +
dy\right) \wedge \sigma_6 \ . 
\label{planesix}
\eeq
The solution presented in equation (26) of \cite{GiHa} reads
\beq ds^2 & =& \left(1 - {2 m \over r^2} \right)   d\tilde{t}^2 +
d\tilde{y}^2 - 2   j
r^2   \sigma_{6}   (d\tilde{t} - d\tilde{y}) - 2 j^2 m \,
r^2 \sigma_{6}^2 \cr &&
+ \left(1 - {2m (1 - 8 j^2 m) \over r^2}\right)^{-1} dr^2 + r^2
d\Omega_3^2
\label{oldsol}
\eeq
Here, we relabeled the coordinates $(t,y)$ of \cite{GiHa} by
$(\tilde{t},\tilde{y})$ to distinguish from the coordinates used in
(\ref{planesix}).

To facilitate the comparison one must rewrite the metric
(\ref{planesix}) in the coordinate chart in which (\ref{oldsol}) is
written.  This is achieved by first twisting
\be
\sigma_{6} \;\; \rightarrow \;\; \sigma_{6} + 2  \beta (dt + dy) \ ,
\ee
and then performing a change of variables
\beq
t & = & (1 + \beta^2  M) \, \tilde t \cr
y & = & -\beta^2  M  \tilde t - \tilde y
\label{coodtrans}
\eeq
so that the metric becomes
\beq
ds_{str}^2 &=& -
\left({1 - {M(1 + \beta^2 M) \over r^2} }\right) \, d\tilde t^2 
+ d\tilde y^2 + \left({\beta \over 1 +
\beta^2 M}\right) \, r^2 \, (d\tilde t - d\tilde y) \, \sigma_{6} \cr
&&  \qquad + {dr^2 \over 1 -{M \over r^2} } + r^2
\, d\Omega_3^2 - {r^2
\over 4} \, \left({\beta^2 M  \over 1 + \beta^2 M} \right)
\, \sigma_{6}^2 \ .  \label{planesix3}
\eeq
Now, (\ref{planesix3}) and (\ref{oldsol}) have the same form, and
they can be seen to be identical if we identify the parameters
according to
\beq
2 m & =&  M   ( 1 + \beta^2 M) \cr
2 j & = & {\beta \over (1 + \beta^2 M) } \ .
\label{paramap}
\eeq
So the dimensionless combination $8  j^2 m$ is related to the
dimensionless combination $\beta^2 M$ by
\be 8 j^2 m = {\beta^2 M \over 1 + \beta^2 M} \ .\ee
The critical value $8 j^2 m = 1$, which causes
the solution (\ref{oldsol}) to degenerate, corresponds to $\beta^2 M$
being infinite.

The fact that (\ref{planesix}) do not asymptote to plane wave geometry
makes the identification of quantities such as mass, entropy, and
temperature even more subtle.  One can nonetheless map
(\ref{planesix}) to Einstein frame and compute the horizon area and
the surface gravity using the killing vector (\ref{killingV}) for
these coordinates.  This yields a $\beta$ independent answer
\be {\cal A}_H = L M^{3/2} \Omega_3 V_{T^4}, \qquad \kappa^2 = {1 \over M}\ . \ee
This was essentially guaranteed based on the arguments presented in
the appendix, and suggests that the parameter $M$ can be interpreted as
a physical mass.

\section{Discussion}

In this article, we presented a powerful technique for generating a
large class of asymptotically plane wave geometries.  Using this
technique, we have succeeded in constructing a supergravity solution
which contains an event horizon while asymptoting to the plane wave
geometry.

With explicit solutions at hand, one can explore various thermodynamic
interpretations of the black string geometry. We find that the null
Melvin twist leaves the area of the event horizon invariant,
suggesting that the entropy of the black string in the plane wave
geometry is identical to the entropy of the black string before taking
the null Melvin twist. Furthermore, with certain assumptions regarding
the definition of temperature in asymptotic plane wave geometries, we
found that it too is left unchanged under the null Melvin twist. This
suggests that the mass parameter of the black string solution
corresponds to a physical measure of mass in some canonical way. It
would be interesting to make these ideas more precise by formulating a
satisfactory definition of mass and temperature in an asymptotically
plane wave geometry.

We also constructed a black string solution in an asymptotically six
dimensional plane wave background using the null Melvin twist
technique.  We showed that this solution is equivalent to the solution
presented in \cite{GiHa}, provided that we map the parameters $(m,j)$
characterizing the mass and the characteristic scale of the plane wave
in \cite{GiHa} to $(M,\beta)$ using (\ref{paramap}). For fixed $j$,
area of the horizon is not monotonic in $m$, whereas for fixed
$\beta$, the area is monotonic in $M$. It is therefore more natural to
define the microcanonical ensemble as fixing $\beta$ instead of fixing
$j$.

Clearly, this null Melvin twist can be used to generate a more general
class of solutions than the ones considered in this article.
Unfortunately, this technique can only be used to construct certain
subset of asymptotically plane wave geometries.  A plane wave geometry
of particular interest which can not be constructed using this
technique is the one which arises as a Penrose limit of $AdS_5 \times
\S^5$.  It would be extremely interesting to find the analogous black
string solution in this asymptotic geometry.

With very little effort, one can dualize the Schwarzschild black
string solution in ${\CP_d}$ to a black hole solution in
$(d-1)$-dimensional G\"odel universe ${\cal G}_{d-1}$.  The horizon
and the velocity of light surface meet only when $\beta^2 M$ is
infinite.

So far, we have only considered black string solutions. It would be
extremely interesting to find the solution corresponding to black
holes. Unfortunately, the null Melvin twist technique involves
T-duality, and can not be applied at the level of supergravity to
backgrounds which do not contain at least one translational isometry
direction. However, since certain plane wave backgrounds can be
constructed using only fields in the NS-NS sector, it may be possible
to extract such a geometry from the consideration of the sigma model
for the string world sheet in this background. A related discussion in
the context of duality between Kaluza-Klein monopole and the NS5-brane
can be found in \cite{Tong:2002rq}.

%%%%%%%%%%%%%%%%%%%%%%%%%%%%%%%%%%%%%%%%%%%%%%%%%%%%%%%%%%%%%%%%%%%%%%%%%%%
\section*{Acknowledgements}
%%%%%%%%%%%%%%%%%%%%%%%%%%%%%%%%%%%%%%%%%%%%%%%%%%%%%%%%%%%%%%%%%%%%%%%%%%%%
We would like to thank
D.~Berenstein,
E.~Boyda,
S.~Ganguli,
O.~Ganor,
P.~Ho\v{r}ava,
G.~Horowitz, and
S.~Ross
for discussions. The work of  EGG is supported
by NSF grant PHY-0070928 and by Frank and Peggy Taplin.
AH is supported in part by DOE grant
DE-FG02-90ER40542 and the Marvin L.~Goldberger Fund.
VEH is supported by NSF Grant PHY-9870115.
OL is supported by NSF grant PHY-0070928.
MR acknowledges support
from the Berkeley Center for Theoretical Physics and also partial support
from the DOE grant DE-AC03-76SF00098 and the NSF grant PHY-0098840.

\section*{Appendix A: Null Melvin twist and the area of the horizon}
         {\setcounter{section}{1} \gdef\thesection{\Alph{section}}}
          {\setcounter{equation}{0}}

In this appendix, we will show that the area of the horizon does not
change under rather general set of manipulations for which the null
Melvin twist is a special case. The proof will be based on an
assumption that the $B_{\mu \nu} $ field for the initial configuration
is regular at the horizon.

Consider starting from a space-time with a horizon and a space-like
translation symmetry along a coordinate $y$ in type II supergravity
theory.  We will assume that the metric\footnote{In this appendix, we
always write the metric in the Einstein frame.} is written in the
Boyer-Lindquist coordinates so that there is a coordinate $r$ which does 
not mix with other coordinates, allowing one to write the metric in
the form
\be g_{\mu \nu} = \left(\begin{array}{c|ccc} g_{rr} & &0& \\  \hline  \\ 0&  & M_{\mu \nu} \\ & \end{array}\right) \ . \ee
One of the coordinates of $M_{\mu \nu}$ will be $t$.  By horizon, we
will mean a fixed $(t,r)$ surface where $g_{rr}$ diverges and $\det M$
vanishes.  Let $A^{\mu \nu}$ denote the cofactor of $M_{\mu \nu}$.
The coordinates are regular away from the horizon, then one can write
\be A^{\mu \nu} = g^{\mu \nu} \det M \ . \label{AasG} \ee
The area of the horizon is the square root of $A^{tt}$ evaluated at
the horizon.

Consider starting from a generic space-time with the properties described 
above,
and applying the following sequence of manipulations:
\begin{enumerate}
\item Boost the space-time along $y$ by $\gamma$
\item T-dualize along the $y$ coordinate
\item Twist by making a change of coordinates
\be x^\mu \rightarrow x^\mu + \alpha^\mu y \ee
\item T-dualize along $y$
\item Boost along $y$ by $-\gamma$.
\end{enumerate}

These manipulations give rise to a new geometry  again in Boyer-Lindquist 
coordinates,   whose horizon area is given by  the square root of 
\beq\label{ProveThis}
A^{tt}+2\sinh^2\gamma\,\alpha^\mu B_{\mu\nu}A^{y\nu}+
\sinh^2\gamma\,\alpha^\mu\alpha^\nu
(B_{\mu\rho}B_{\nu\sigma}+g_{\mu\rho}g_{\nu\sigma})A^{\rho\sigma}
\eeq Using the relation (\ref{AasG}) away from the horizon, we can
rewrite the last term in this expression as
\beq
\sinh^2\gamma\,\alpha^\mu\alpha^\nu
(B_{\mu\rho}B_{\nu\sigma}+g_{\mu\rho}g_{\nu\sigma})A^{\rho\sigma}=
\sinh^2\gamma\,~ {\mbox{det}}M\, \alpha^\mu\alpha^\nu
(B_{\mu\rho}{B_\nu}^\rho+g_{\mu\nu})
\eeq
This is a scalar multiplied by ${\mbox{det}}M$, so it has to
vanish at the horizon. The second term is an expression proportional to 
\be V_\mu A^{\mu\nu}=V^\nu{\mbox{det}}M \label{gotozero} \ee
if we set $V_\mu = \alpha^{\nu} B_{\mu \nu}$ which we assume to be
regular at the horizon. To prove the invariance of the area, it is
sufficient to prove that for any regular vector $V_\mu$, 
(\ref{gotozero}) goes to zero as we approach the horizon.

To show this we introduce a vielbein ${e_\mu}^a$ for the $D-1$
dimensional ``metric'' $M_{\mu\nu}$: 
\be
M_{\mu\nu}=\eta_{ab}{e_\mu}^a{e_\nu}^b.
\ee 
Since there is no
curvature singularity at the horizon, all components of the vector
$V^a$ must be regular there. In terms of the vielbein ${e_\mu}^a$ and its cofactor ${E^\mu}_a$, we can write
\be
V_\mu A^{\mu\nu}=-{e^\nu}_a V^a({\mbox{det}}~e)^2=-{E^\nu}_a V^a
{\mbox{det}}~e.  \ee 
Since all elements of $M_{\mu\nu}$ are finite at the horizon, we can
choose all ${e_\nu}^a$ to be finite as well. This in turn leads to
finite minors ${E^\nu}_a$. On the other hand, since ${\mbox{det}}M$
vanishes at the horizon, so does ${\mbox{det}}~e$, proving that $V_\mu
A^{\mu\nu}$ vanishes at the horizon for any value of $\nu$.

This shows that the area of the horizon is invariant with respect to
sequence of manipulations described in this appendix. In particular,
by taking $\alpha^\mu$ to rotate the four independent transverse
rotations of $\S^7$ and scaling the twist with respect to the boost,
we show that the area of the horizon is invariant under null Melvin
twist.

%%%%%%%%%%%%%%%%%%%%%%%%%%%%%%%%%%%%%%%%%%%%%%%%%%%%%%%%%%%%%%%%%%%%%%%%%%%%%%%%%%%%%%%%%%%%%%%%%%

\providecommand{\href}[2]{#2}\begingroup\raggedright\endgroup

\end{document}